# A Chessboard Model of Human Brain and One Application on Memory Capacity

Chenxia Gu#, Shaotong Wang#, Hao Yu*

(Affiliation): Department of Mathematical Sciences, Xi'an Jiaotong-Liverpool University, Suzhou, 215123, China

#These authors contributed equally to this work.
*Corresponding author Email: hao.yu@xjtlu.edu.cn

## Abstract

The famous claim that we only use about 10% of the brain capacity has recently been challenged. Researchers argue that we are likely to use the whole brain, against the 10% claim. Some evidence and results from relevant studies and experiments related to memory in the field of neuroscience leads to the conclusion that if the rest 90% of the brain is not used, then many neural pathways would degenerate. What is memory? How does the brain function? What would be the limit of memory capacity? This article provides a model established upon the physiological and neurological characteristics of the human brain, which could give some theoretical support and scientific explanation to explain some phenomena. It may not only have theoretically significance in neuroscience, but could also be practically useful to fill in the gap between the natural and machine intelligence.

## 1. Introduction

The memory of a human being is far more complex than storing and retrieving stored information in a computer. It includes the association and inference [1, 2] among information, e.g., the association between two pieces of information such as the appearance of an apple and the human-language name, apple. To store the association, neural pathways will be generated, i.e., connections, will be made and more neuron cells must be involved than to simply memorize the two pieces of information [3]. The nerve cells can perform two states excitatory and inhibitory. The transformation of a nerve cell between the two states depends on the irritation obtained from the external environment [4] [5]. The memory forms and stores in hippocampus which consists of a large number of nerve cells. Every pair of nerve cells can generate a connection named synapse which is a channel for transmitting the nervous impulse from one cell to another. In addition, synapse can only conduct impulse in one direction, which means impulse cannot transmit back to the original cell through the same synapse [6]. In terms of the formation delivered by synapse, the excitatory nerve cells have a probability to establish synapse with adjacent cells. The excitement tends to propagate from the cells connected with excitatory sensor cells to the target neuron cells. Whenever the neuron cell state changes, the appeared synapse can always be used as a channel for conducting excitement. The pattern of connections among cells is regarded as the memory of creatures [7].

Inspired by biological neural networks, artificial neural networks (ANNs) are a family of statistical learning models used to estimate or approximate unknown functions that can depend on a large number of inputs, which are presented as systems of interconnected "neurons" which exchange messages between each other [1]. Inheriting the structure of ANNs and its application on object recognition, a model is established to explain the functioning of the brain, based on the physiological and neurological characteristics of the human brain. Binary number will be used to describe the state of one neuron cell where 1 means excitatory and 0 means inhibitory. The two states of a cell can change when the condition is just right [8]. Then the bridges of each pair of nerve cells represent the synapse. Finally, the outermost layer of the processor matrix represents the target cells which are the end of excitement propagation [9].

## 2. Modelling

The model consists of a sensor-matrix layer and a processor-matrices layer, each of which contains a certain number of sensor-matrices and processor-matrices respectively. The reason why we can work within two-dimension matrix is that the exact positions of neuron cells are not so important, since essentially speaking, memory depends on not the neurons themselves, but the connections between them. Moreover, due to the complexity of the real process in the human brain which could be far beyond our imagination, not even within three dimensions could it be fully illustrated [10-12].

Sensors are used to generate perception which responds to a stimulation from the outside world. For instance, our eyes, as a visual sensor, can sense the visual features of the stimulation; ears can sense vocal information and hands can sense tactual information. All the information will then be processed and transmitted to the processors, where memory-making process takes place (hippocampus) [13]. After the processors received the information from the sensors, the different sorts of information will be associated together in some ways and forms an initial acquisition of the memory subject. The whole memory process involves three steps, Encoding, Consolidation and Storage, and Retrieval [14] [15]. The transition process from sensors to processors is called encoding step. The association of all sorts of information that the processors received from sensors is regarded as the step for consolidation and storage. For instance, if the visual features of an apple and the literal information of an apple are associated in a memory process, then next time, one could recall the name, apple, as soon as he or she see an apple. The recalling is viewed related to the retrieval step.

Although this model does not require a particular memory representation, it is useful to illustrate the realization of all the three steps with a simplified representation.

### 2.1. Encoding

To simulate transition process, the system with sensors and processors is considered. Each state of the sensors or processors can be stored in a matrix, where each entry represents the state of each neuron cell. The state of a neuron cell is assumed to be either excitatory or inhibitory, which is indicated by binary numbers, that is, 1 and 0, respectively. An example of the encoding process from the sensors to the processors is illustrated in **Figure 1.**

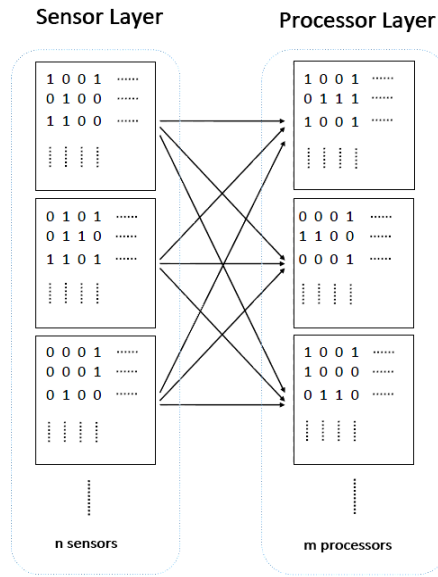

Figure 1 n to m correspondence: an excitatory cell is denoted by the number 1.

Every cell in the sensor matrix could have its own multiple corresponding cells in the processor matrix. If one cell in the sensor matrix is set to be 1, then each of the corresponding cells in the processor matrix will have the value 1 as well [16]. In this model, the correspondence is set to be one-to-one and is randomly generated. The randomness of correspondence is of computational significance, since it disperses the distribution of the cells with value of 1 in the processor which maximizes the utilization of the matrix and maximizes the distinctions of the information in the sensor. Imagine if the cells with value 1 in the processor are all concentrating in the center, the results from the operation of the model will be hard to distinguish.

## 2.2. Consolidation and Storage

Now constrain our attention to one of the sensors and its corresponding processor, say visual sensor and its corresponding processor, and the correspondence between them is assumed to be one-to-one. Therefore, the visual processor, which has a spatial structure in the brain, could be simply represented by another n by n matrix. Because the message delivered fast in human brain, the all duration of the transmission from first cell to another cell are regarded equally [17]. At first, the processor received an initial state of excitation from the sensor. Then, the excitation transmission process starts. (It is assumed that the excitation could only conduct from one cell to another once in one stage of the process and it could only be conducted from the newly excited cells. In another word, the cells that have successfully conducted excitation will not be able to conduct excitation any more. After the transmission from first cell to another, a new state of this matrix will be generated.) Before the conduction of excitation, the path which used to transmit the excitation between a pair of cell is generated firstly. When the conduction process is finished, the path will be remained there, which forms the memory. The term bridge is used to indicate whether there is path from one cell to another, and noticed that every bridge is represented by a directed line segment. If there is a bridge from cell A to cell B, then it means excitation can be conducted from cell A to cell B. A pair of cells could have at most two bridges between them, opposite in direction. For an illustration, the cues of the n by n processor matrix are assumed to lie on the four edges excluding the points at the corners, as is shown in Figure 2. The inner matrix represents the processor with some initial state and the red cells on the peripheral edges around this matrix represent the cues.

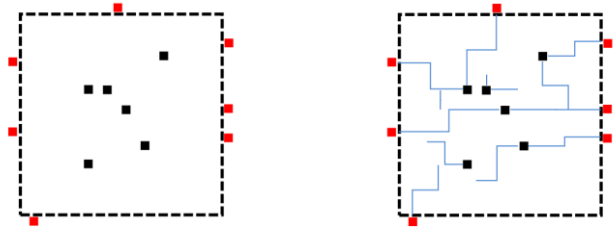

**Figure 2 .** Initial configuration and one possible pathway

The conduction process in a processor is regarded as finished once all the cues received the excitation transmitted by bridge and series of excitation cell. The transmission of the excitation is under some conduction rules with neurological significance and it is obvious that the pathway of the excitation will be governed by the cues. Once the initial configuration and the cues are determined, the excitation will be under conduction between cells in the processor and ends up with its reach to the cues. One possible pathway is also shown in **Figure 2** where the bridges are indicated by the blue line segments; red cells represent cues.

### 2.3. Conduction Rules

To deduct the possible pathway, conduction rules are established to ensure that all the cues could be reached and the establishment of the bridge is optimized. For better illustration, several rules have been declared. Firstly, the conduction only occurs when bridge has been established between the two cells. Secondly, design that each cell can build up bridges to its four adjacent cells, with directions left, right, up and down respectively. The circumstance with more connections will be considered in later analysis. Thirdly, the transmission process is assumed to be inherently probabilistic. The distance to a cue determines the probability whether the bridge will be established between the cell and the next cell which is nearer to that cue.

Let O and I denote the sets of cues and original excited cell respectively. Suppose that there are totally m cues and n original excited cells. That is to say, $O = O_1, O_2, O_3, \ldots, O_m$ and $I = I_1, I_2, I_3, \ldots, I_n$. Then theoretically, the conditional $f_{O_i I_3} = prob(O_i|I_j)$ is determined by

$$f_{O_i I_j} = \frac{P_{O_i I_j}}{\sum_{k \in O} P_{O_k I_j}} \qquad (1)$$

Where $P$ denotes $prob(O_r \cup I_t)$, the probability that events $O_r$ and $I_t$ happen simultaneously. Obviously, by assumption, the instantaneous set of unreached cues governs the conduction of the next moment. Moreover, the cues could have accumulative effects on the probability of the bridge establishment. After the transmission process, the bridges are kept and stored in the processor, which is regarded as memory.

In the second time, when the figure of apple irritates the sensor again, the excitement will be transmitted automatically through the path established in the first time and all cues will be irritated again which indicates the name, apple.

### 2.4. Illustration

The model is aimed to simulate the memory process and explore the memory capacity. In another word, it is aimed to find out the limit beyond which the brain would be overloaded and the association of different types of information could be unreachable. For better illustrate, the model is now applied to a specific example with a 30*30 processor matrix. The initial configuration is initialized as shown in **Figure 3**.

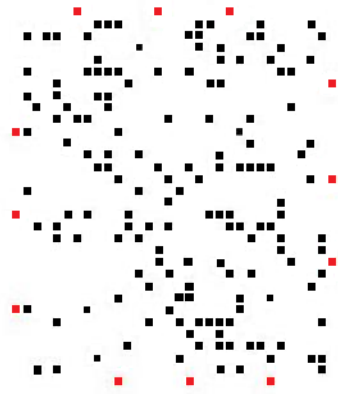

**Figure 3.** Initial configuration with black dots indicating the cells in the processor and red dots indicating the cues.

The black dots inside represents the initial state of the processor, which could be an image of an apple. If the red dots outermost refer to the name, apple, then after application of the model, bridges will be established through which the excitation will be conducted until the reach of excitation to all the cues eventually. One possible bridges is shown in **Figure 4.**

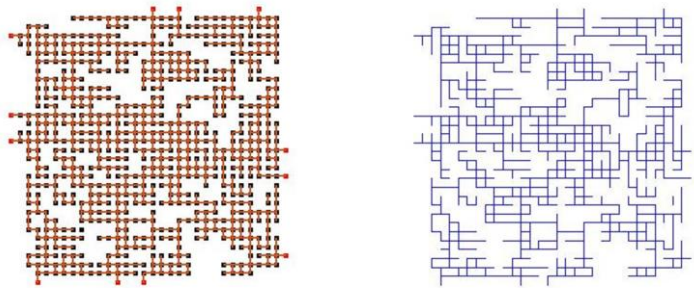

**Figure 4.** One possible bridges.

After that, the bridges will be kept as a memory which associates the vision of the apple and its name. Then next time, if the processor receives the information of the apple again, the excitation will be transmitted along the bridges stored automatically and finally reach the cues which indicates the name of the apple. If the processor receives a different piece of information, the excitation will still be transmitted but may not be associated with the same cues, the name apple. For instance, if an image and its name (cues represented by red cells) are defined as shown in **Figure 5 (a),** another different image indicated in **Figure 5 (b)** input in a short time. Then the name output in **Figure 5 (a)** is different from it in **Figure 5 (b)**, so this is a successful case. However in **Figure 5 (c)**, another different image obtained by processor in a short time, the name output is totally the same as it in **Figure 5 (a)**, which means the images are not be distinguished. Therefore this is a chaotic case [18].

(a)

(b)

(c)

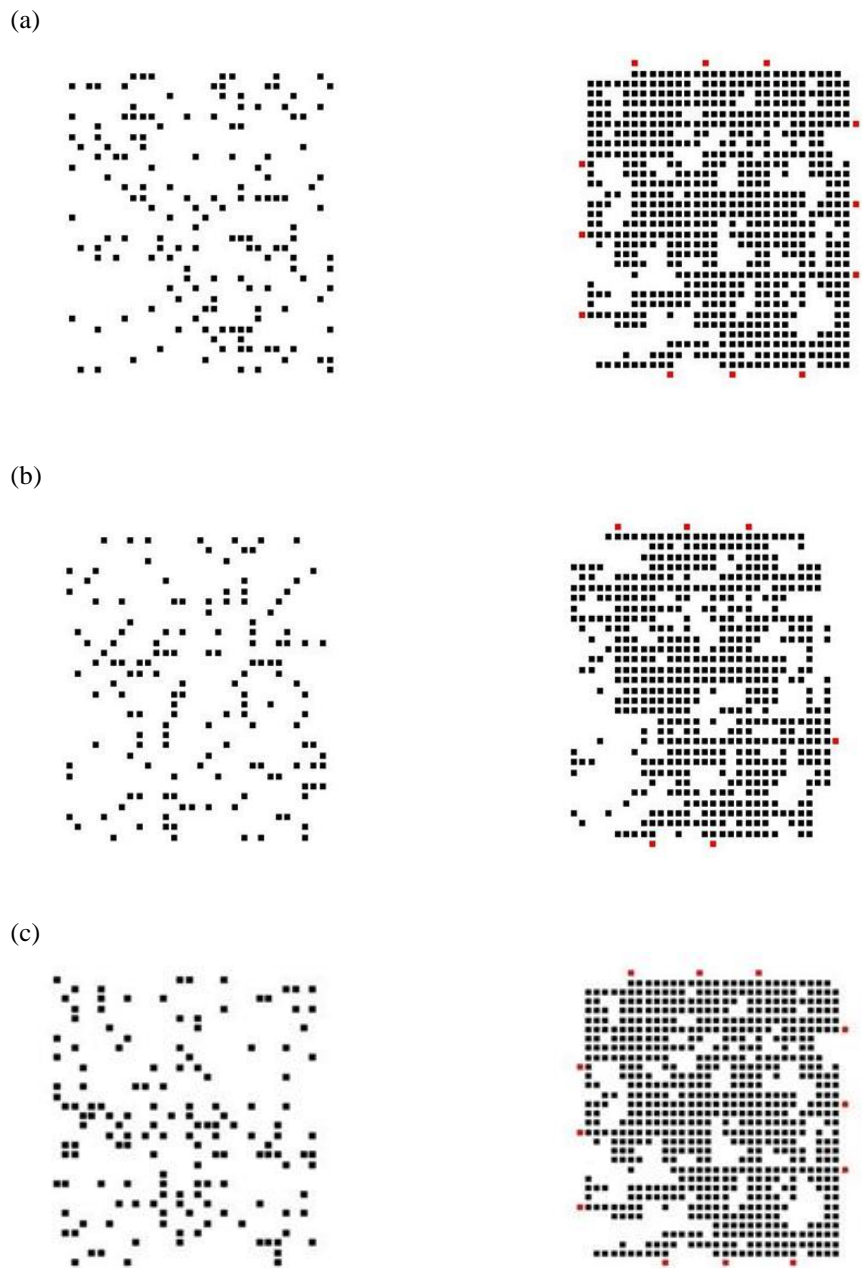

**Figure 5** (a)The reference image and its name. (b)Successfully-Distinguished Case: the processor successfully distinguishes a different image. (c)Chaotic Case: the associated cues are exactly the same as shown in (a), while the information that the processor rece

## 3. Results and Analysis

Whether the system is chaotic depends on the state of the cells [19]. In order to describe the chaotic degree, the concept similarity is introduced here. Two pieces of different image information are delivered to the same processor in a short period, which means when the second image message is obtained, the bridges established for the first image have not degraded. Two cues, one is for the first message and the other is for the second message. Some parts in the two cue-layers generated respectively are supposed to be different, and the percentage of the same part in the two cue-layers is indicated by similarity. However if the two cue-layers are exactly the same, the case is regarded as a chaotic case. First of all, assume that the number of cells with value of 1 in inner matrix represents the number of information cells included in the image. The relationship between whether the processor is chaotic and the number of information cells involved is investigated by the following method. First, change the percentage of the number of cell with the value of 1 among total number of cells both in image matrix and cue layer from 1% to 90%. Then, repeat each test 20 times for each percentage and record the similarity of each different test. The similarity (S) of the results can be calculated by:

$$G_{ij} = \begin{cases} 0 & if\ i \neq j \\ 1 & if\ i = j \end{cases}, \quad S = \frac{\sum_{t=1}^{n} G_{a_t b_t}}{n} \times 100\% \tag{2}$$

Where n is the total number of bits in cue layer; $a_t$ and $b_t$ represent the $t^{th}$ bit in the test sample and the original sample, respectively.

When S equals 1, which means the model cannot tell the difference between the two different images, which is called chaotic. If G is 0, that means the two relative cues are not exactly the same, the difference of the two different images is told by the model. In order to find an optimal condition to distinguish the different image information, first of all, assume that the number of cells with value of 1 in image matrix represents the amount of information stored in the image matrix. The relationship between the whether the processor is chaotic and the number of cells is investigated by using the following method. Adjust the percentage of the cells storing information and located in image matrix and cue layer from 1% to 90%. Repeat 50 times in every percentage and record the times where the similarity is not equal to 1. Then calculate the successful rate under each condition, where the successful rate is the percentage of the experiment with S=1 in the whole 50 times experiment. Therefore, Table 1 shows the results.

Table 1 Successful Rate

| Cue Layer / Internal Matrix | 10% | 30% | 50% | 70% |
|---|---|---|---|---|
| 1% | 86% | 68% | 46% | 30% |
| 5% | 96% | 86% | 74% | 70% |
| 10% | 98% | 82% | 88% | 76% |
| 20% | 90% | 82% | 90% | 68% |
| 30% | 86% | 84% | 72% | 74% |
| 40% | 74% | 76% | 70% | 72% |
| 50% | 74% | 76% | 68% | 68% |
| 60% | 66% | 68% | 70% | 64% |
| 70% | 58% | 70% | 52% | 70% |
| 80% | 60% | 64% | 58% | 62% |
| 90% | 56% | 44% | 56% | 56% |

Based on Table 1, a fitting surface is plotted to find the optimal condition, shown in Figure 6.

The top point of the curve in Figure 6 with the highest success rate (98%) is the highest point in the whole figure. The condition where the percentage of cells storing information in cue-layer is 10% and the one in image matrix is also 10% is the optimal condition. Therefore, the test result is record which is in condition 10% density image matrix and 10% cue fulfillment. That means when the 10% of the brain space is used to store information, the effect of remembering association is best. That is why in human beings brain, 10% of the storing space is occupied. However, other cells which not hold information are not useless and they are used to transfer excitation. In the course of human evolution, the people with a poor memory function are eliminated and the people with brain in the optimal condition are reserved.

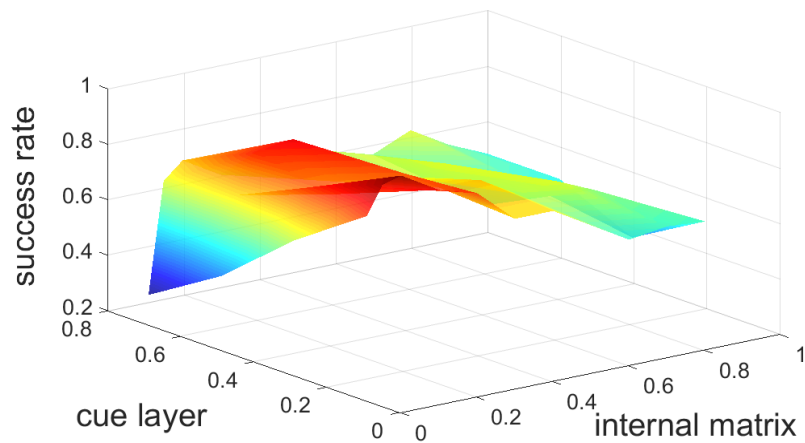

**Figure 6.** Fitting surface of successful rate.

Image what would happen if much more storage space is used or unbalanced storage space is used. If 70% of storage space is used both of character part and image part, the successful rate only 70%. That means if you told the patients that image represent an apple, in the next time he may call the other totally different image an apple [20]. Therefore the patients will be diagnosed for a psychopath whose memory is overloaded.

Moreover, in another condition where the two percentages of storage space of character part and image part are different, the function of brain is also influenced. For instance when the character capacity is 10% and the image part is 70%, the success rate is only 58%. That means if the man who has great talent in memorizing images but his capacity of character memorizing is not balanced with it of image memorizing, the man is also has a tendency to be insane. In other words, the genius who has gift on one aspect is nearly lunatic.

Generally, this model can be regarded as a mini-brain. The ability of study depends on the number of cells in use, which means that the length of the period for study decreases with the size of the matrix [21]. The biggest differences between this model and the real brain are the number of cells and the number of connections. Therefore, in this section, matrices with different dimensions and different numbers of connections are adjusted and then the results with different parameters are compared.

Firstly, the dimension of the matrix is expanded to 50×50, 80×80 and 100×100, respectively. Thus, correspondingly, the number of neural cells involved increases to 2500, 6400 and 10000, respectively. The condition with 10% density image matrix and 10% cue fulfillment (the optimal condition) are set to examine the effect of the size of the matrix on the successful rate. After that the case with dimension 50×50 has been repeated 50 times and a new successful rate 98% was obtained which is also comparatively high. The cases with dimension 80×80 and 100×100 were tested and the successful rates are 100% and 98%, respectively. Therefore, we can speculate that the successful rate may be independent from the size of matrix.

Secondly, in order to investigate whether the number of connections can affect the results of the experiment, we double the number of connections and observe, i.e., not only the up, below, left and right adjacent cell are connected to the central cell but also the up-right, up-left, bottom-left and bottom-right adjacent cell can be associated. Two pieces of different image information are given under the optimal condition. The relative cue-layers are different so the two pieces of image are successfully distinguished.

After that, the experiment has been repeated 50 times and the successful rate obtained is 100%. Therefore we can speculate that the number of connections may not influence the successful rate either. According to the statistics, the dimension and the number of connections may have no effect on the successful rate. Therefore, we can conclude that the percentage of optimal occupied space in the brain is about 10%.

Finally, to simplify the computation and programming, the number of connections that a neuron cell could have with the adjacent cells is up to eight in this model. However, in fact, thousands of connections could be made by one single cell [21].

Hence, further study could enlarge the matrix. Since in the model the matrix used to represent the processor has comparatively small dimensions, the processor is considered as a whole. When the matrix becomes larger and larger, partition of the processor could be taken into consideration.

In this model, only two states (excitatory and inhibitory) of a cell are considered. More values could be used to represent different states to increase the complexity of the situation. Moreover, the transmission is assumed to happen so quickly that forgetting can be neglected. In some other cases, forgetting may have significant effects on the memory process.

## 4. Conclusion

The model has been established to simulate the memory process of the human brain. Through conducting thousands of experiment by simulation, the statistical results indicate that the optimal percentage of occupation in the brain is 10%. However, the other cells which are not occupied are necessary for transmitting the excitation rather than unexploited. Furthermore, it can be speculated that the successful rate may be independent from the size of the matrix. And the dimension and number of connections may have no effect on the successful either.

## Acknowledgements

This work was supported by grants from the National Natural Science Foundation of China (No.11204245), and the Natural Science Foundation of Jiangsu Province (No.BK2012637).